\documentclass[a4paper,11pt]{article}
\usepackage{siunitx}
\usepackage{amsmath}
\usepackage{jinstpub}
\usepackage{lineno}
\usepackage{comment}

\title{\boldmath 
A Novel Nuclear Emulsion Detector for Measurement of Quantum States of Ultracold Neutrons in the Earth's Gravitational Field}

\author[1]{Naoto~Muto} \author[2]{Hartmut~Abele} \author[3]{Tomoko~Ariga} \author[2,4]{Joachim~Bosina} \author[5]{Masahiro~Hino} \author[1,6]{Katsuya~Hirota} \author[6,7]{Go~Ichikawa} \author[4]{Tobias~Jenke} \author[1]{Hiroaki~Kawahara} \author[6]{Shinsuke~Kawasaki} \author[1,8]{Masaaki~Kitaguchi} \author[2,4]{Jakob~Micko} \author[6,7]{Kenji~Mishima} \author[1,9]{Naotaka~Naganawa} \author[1,8,9]{Mitsuhiro~Nakamura} \author[4]{St\'ephanie~Roccia} \author[1,9]{Osamu~Sato} \author[2]{Ren\'e~I.~P.~Sedmik} \author[10]{Yoshichika~Seki} \author[1,6]{Hirohiko~M.~Shimizu} \author[1]{Satomi~Tada} \author[1,11]{Atsuhiro~Umemoto}
\affiliation[1]{Department of Physics, Nagoya University,\\
Furo-cho, Chikusa-ku, Nagoya, 464-8601, Japan; muto@flab.phys.nagoya-u.ac.jp,\\ hirota@phi.phys.nagoya-u.ac.jp, kawahara@flab.phys.nagoya-u.ac.jp, kitaguchi@phi.phys.nagoya-u.ac.jp,  naganawa@flab.phys.nagoya-u.ac.jp, nakamura@flab.phys.nagoya-u.ac.jp, sato@flab.phys.nagoya-u.ac.jp, shimizu@phi.phys.nagoya-u.ac.jp, tada@flab.phys.nagoya-u.ac.jp}
\affiliation[2]{Atominstitut, Technische Universit\"at Wien,\\
Stadionallee 2, 1020 Vienna, Austria; hartmut.abele@tuwien.ac.at, joachim.bosina@tuwien.ac.at,\\ rene.sedmik@tuwien.ac.at}
\affiliation[3]{Kyushu University,\\
744 Motooka Nishi-ku, Fukuoka, 819-0395, Japan; tomoko.ariga@cern.ch}
\affiliation[4]{Institut Laue-Langevin,\\
71 avenue des Martyrs CS 20156, 38042 GRENOBLE Cedex 9, France; jenke@ill.fr, micko@ill.fr, \\roccia@ill.fr}
\affiliation[5]{Institute for Integrated Radiation and Nuclear Science, Kyoto University,\\
Kumatori, Osaka, 590-0494, Japan; hino.masahiro.2x@kyoto-u.ac.jp}
\affiliation[6]{High Energy Accelerator Research Organization (KEK),\\
Tsukuba, Ibaraki, 305-0801, Japan; go.ichikawa@kek.jp, shinsuke.kawasaki@kek.jp, kenji.mishima@kek.jp}
\affiliation[7]{Japan Proton Accelerator Research Complex (J-PARC) Center,\\
Tokai, Ibaraki, 319-1195, Japan}
\affiliation[8]{Kobayashi-Maskawa Institute for Origin of Particles and the Universe (KMI), Nagoya University,\\
Furo-cho, Chikusa-ku, Nagoya, 464-8601, Japan}
\affiliation[9]{Institute of Materials and Systems for Sustainability, Nagoya University,\\
Furo-cho, Chikusa-ku, Nagoya, 464-8601, Japan}
\affiliation[10]{Center for Physics and Mathematics, Osaka Electro-Communication University,\\
Neyagawa, Osaka 572-8530, Japan; seki@osakac.ac.jp}
\affiliation[11]{Division of Physics, University of Tsukuba,\\
Tennodai, Tsukuba 305-8571, Japan; umemoto.atsuhiro.gu@u.tsukuba.ac.jp}

\emailAdd{muto@flab.phys.nagoya-u.ac.jp}

\abstract{
Hypothetical short-range interactions could be detected by measuring the wavefunctions of gravitationally bound ultracold neutrons on a mirror in the Earth's gravitational field.
Searches for them with higher sensitivity require detectors with higher spatial resolution.
We developed and have been improving an ultracold neutron detector with a high spatial resolution, which consists of a Si substrate, a thin converter layer including $^{10}$B$_{4}$C, and a layer of fine-grained nuclear emulsion.
Its resolution was estimated to be less than \SI{100}{\nano \metre} by fitting tracks of either $^{7}$Li nuclei or $\alpha$-particles, which were created when neutrons interacted with the $^{10}$B$_{4}$C layer.
For actual measurements of the spatial distributions, the following two improvements were made. 
The first improvement was to establish a method to align microscopic images with high accuracy within a wide region of \SI{65x0.2}{\milli \metre}.
We created reference marks of \SI{1}{\micro \metre} and \SI{5}{\micro \metre} diameter with an interval of \SI{50}{\micro \metre} and \SI{500}{\micro \metre}, respectively, on the Si substrate by electron beam lithography and realized a position accuracy of less than \SI{30}{\nano \metre}. 
The second improvement was to build a holder for the detector that could maintain the atmospheric pressure around the nuclear emulsion to utilize it under vacuum during exposure to ultracold neutrons.

The intrinsic resolution of the improved detector was estimated to be better than \SI{0.56 \pm 0.08}{\micro \metre} by evaluating the blur of a transmission image of a gadolinium grating taken by cold neutrons.
The evaluation included the precision of the gadolinium grating.
A test exposure was conducted to obtain the spatial distribution of ultracold neutrons in the quantized states on a mirror in the Earth's gravitational field.
The distribution was obtained, fitted with the theoretical curve, and turned out to be reasonable for UCNs in quantized states when we considered a blurring of \SI{6.9}{\micro \metre}.
The blurring was well explained as a result of neutron refraction due to the large surface roughness on the upstream side of the Si substrate.
By using a double-side-polished Si substrate, a resolution of less than \SI{0.56}{\micro \metre} is expected to be achieved for UCNs.
}

\proceeding{December 27th, 2021}
\begin{document}
\maketitle
\flushbottom

\section{Introduction}
\label{sec:intro}
Hypothetical fields manifesting as dark energy or dark matter may also lead to interactions causing effective deviations from Newtonian gravity at short distances.
In recent years, various experiments have been searching for such interactions~\cite{short_for_exp1}.
Some of them have been searching hypothetical fields by measuring the spatial distributions of quantized states of ultracold neutrons (UCNs) bound by the Earth's gravitational potential on mirrors and comparing those distributions to theoretical curves including only Newtonian gravity~\cite{short_for_exp2,short_for_exp3,short_for_exp4,Westphal2007}.
Those distributions have characteristic structures in the scale of ten micrometers, and therefore they require detectors with a high spatial resolution of not more than a few micrometers.
The sensitivity of those experiments depends on the spatial resolution of the detectors and the experimental setup including mirrors.

So far, experiments studying gravitational levels have been using spatial resolution detectors made of CR39, originally proposed in~\cite{short_for_exp2}, or a pixelated Si detector equipped with a Ni-coated cylindrical glass rod to expand the images~\cite{short_for_exp4}. Measurements using CR39 detectors with 235-U-converter are presented in~\cite{Nesvizhevsky2005, Baessler2011a}.
The \textit{q}\textsc{Bounce} collaborations has been using detectors with 10-B-converter~\cite{short_for_exp3}.
Its handling and optical readout is presented in~\cite{jenke2013ultracold}. Both concepts reach a spatial resolution in the order of 1--2 \si{\micro \metre}.
The Si detector with the cylindrical rod reached a resolution of \SI{0.7}{\micro \metre}.

To further increase the sensitivity, a detector with a higher spatial resolution is necessary.
We have developed such a detector based on a fine-grained nuclear emulsion. The resolution obtained is superior by more than one order of magnitude.

Nuclear emulsions consist of silver halide crystals dispersed in gelatin. 
Particles passing through the emulsion leave a linear series of silver grains (a track) that can be detected optically.
The high spatial resolution of these tracks has led to important results in nuclear and particle physics, such as the discovery of $\pi$ mesons~\cite{meson}, studies on double hypernuclei~\cite{double_hyper1,double_hyper2,double_hyper3}, and the first detection of tau neutrino interactions~\cite{tau}.
Recently, a fine-grained nuclear emulsion was developed for use in experiments searching for dark matter~\cite{fine_grained}.
With a crystal size of \SI{40}{\nano \metre} and a density of 13 crystals per \si{\micro \metre}, our emulsion detectors are 6 times denser and have 5 times smaller crystals than a normal emulsion, thereby improving the spatial resolution.
In addition, no track is formed by minimum ionizing particles or electrons produced by $\gamma$-rays occurring as background in neutron experiments because silver halide crystals have small sizes and have not been sensitized.

\begin{figure}[htbp]
\begin{center}
\includegraphics[scale=0.28,clip]{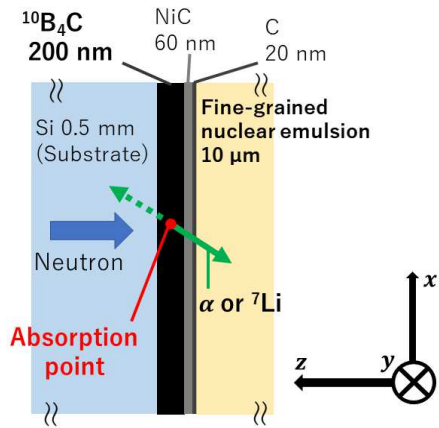}
\caption{Schematic cross-sectional view of our high-resolution nuclear emulsion detector for cold neutrons and UCNs.
Neutrons enter into the detector from the left.
A $^{10}$B$_{4}$C layer converts the neutrons to $\alpha$-particles and $^7$Li particles, emitted back-to-back.
Both of these create a track in the emulsion on the right side.}
\label{fig:nuclear_emulsion}
\end{center}
\end{figure}

\begin{figure}[htbp]
\begin{center}
\includegraphics[scale=0.35,clip]{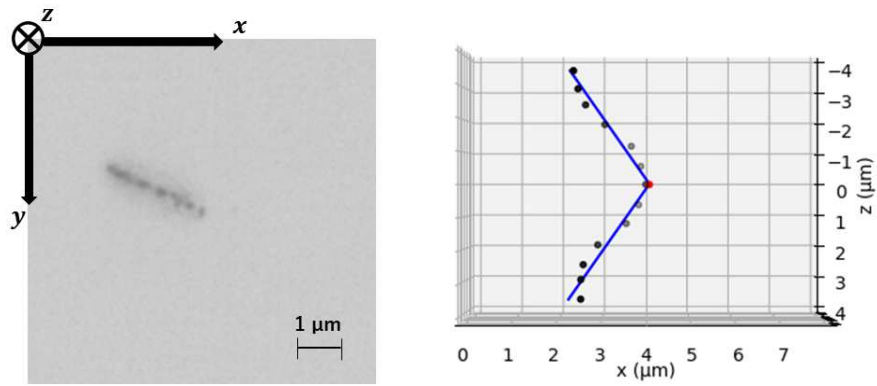}
\caption{Track of an $\alpha$-particle visible in an optical microscope image (left), and the reconstructed track in the $x$-$z$ plane (right).
Pixels with low brightness correspond to silver particles, which are recognized as grains.
Right: Black dots show the grain coordinates reconstructed from the figure on the left.
The blue line shows a reconstructed track obtained from fitting the grains.
The red dot shows the coordinates of the point of the boundary between real and mirror images on the surface of the C layer ($z$ = 0) resulting from the line fits.}
\label{fig:track}
\end{center}
\end{figure}

Emulsion detectors can be sensitive to cold neutrons or UCNs when using a $^{10}$B$_{4}$C conversion layer sputtered onto a Si substrate, followed by NiC, and C layers~\cite{neutron_emulsion}.
The emulsion is applied above the C top layer (Figure~\ref{fig:nuclear_emulsion}).
NiC and C act as stabilizers for $^{10}$B$_{4}$C and the emulsion, respectively.
The thickness of $^{10}$B$_{4}$C, NiC, and C are \SI{200}{\nano \metre}, \SI{60}{\nano \metre}, and \SI{20}{\nano \metre}, respectively.
When neutrons interact with the $^{10}$B$_{4}$C layer, $^{7}$Li nuclei and $\alpha$-particles are emitted back-to-back by neutron capture reactions and either of them passes through the emulsion layer.
The detection efficiency is estimated to be \SI{40}{\percent} for neutron velocities around \SI[per-mode=symbol]{10}{\metre\per\second}.
The emulsion is capable of recording tracks of $^{7}$Li nuclei with a range of \SI{2.7 \pm 0.4}{\micro \metre} or $\alpha$-particles with a range of  \SI{5.2 \pm 0.4}{\micro \metre} in the emulsion, calculated by SRIM-2013 based on the composition of the recent fine-grained nuclear emulsion.
Tracks from the absorption points of neutrons were observed under an optical microscope with an epi-illumination system~\cite{microscope}.
The wavelength of the illumination was \SI{455}{\nano \metre}, and the microscope was equipped with a CMOS image sensor.
We used an oil immersion lens with magnification $100\times$, and a numerical aperture of 1.45.
As a feature of the detector, both real and mirror images of the tracks are obtained because the surface of the C layer in the substrate functions as an optical mirror (Figure~\ref{fig:track}).

The coordinates of the absorption points are obtained by the following eight steps.
First, tomographic images are acquired by taking images at different heights in 63 steps distributed equally over the thickness of the emulsion layer.
Each image has a field of view of \SI{113x113}{\micro \metre} and the resolution is $2048\times2048$, resulting in a pixel size of \SI{55x55}{\nano \metre}.
The thickness of each imaging layer (corresponding to the vertical resolution) is \SI{0.37}{\micro \metre} prior to shrinking of the emulsion during development.
Second, pixels with low brightness values corresponding to silver particles are detected in the acquired images, and those pixels are recognized as a part of grains.
This detection is performed twice: once for images that only contain real images and again for images containing the mirror images of the tracks.
Third, grains in the lowermost 1.5 layers above the base layer are selected.
These grains correspond to the starting points of tracks in the emulsion.
Fourth, pairs of two grains corresponding to the starting and ending point of original tracks whose distance is consistent with the theoretical track length (more than \SI{1.5}{\micro \metre} and less than \SI{7.5}{\micro \metre}) are assigned.
If the density of original tracks is high, the starting or ending points of two or more tracks may be mixed during the selection process.
In this case, there are no silver particles on the three-dimensional straight line connecting the starting and ending points, which means that the total brightness value of the pixels on this line is higher than it should be for a real track.
Fifth, to reject these pairs, if the distance between the starting points or ending points of different pairs is less than \SI{0.8}{\micro \metre}, the pairs with the higher total brightness value are excluded.
Sixth, for the remaining starting/ending point pairs having proper distances, tracks are reconstructed from the grains lying on the connecting line between the two points.
These are called subsequently `reconstructed tracks'.
Seventh, the coordinates of the point of the boundary between real and mirror images at the surface of the C layer, and the slopes of each reconstructed track are obtained by fitting both images of each track with symmetrical lines.
Finally, absorption points of neutrons are identified by extrapolating the reconstructed track of the real images to the middle of $^{10}$B$_{4}$C layer using the fitting results.
The spatial resolution of this detector was calculated from the data of Reference~\cite{neutron_emulsion} to be less than \SI{100}{\nano \metre}.
Note that we only consider tracks with angles $\theta \leq $ \SI{0.9}{\radian} between the track and the normal line to the surface of the C layer, as for tracks at larger angles, the error increases significantly.

In order to utilize the detector in the measurement of the spatial distribution of UCNs, two key issues had to be resolved.
The first issue was to establish a method to ensure that high spatial accuracy and resolution in a detection area of \SI{65x0.2}{\milli\metre}, where a field of view of the microscope is \SI{113x113}{\micro \metre}.
This cannot be guaranteed by simply stitching images from different positions, as offsets accumulate and lead to increasing errors.
Therefore, we created microscopic marks on the Si substrate that serve as independent reference points to align the optical images.
We demonstrated this method to work and the achievement of high spatial resolution by an analysis of the transmission image of a gadolinium grating taken by cold neutrons at the Japan Proton Accelerator Research Complex (J-PARC).
The second issue was to find a method allowing us to utilize the emulsion detector in a vacuum of \SI{1}{\pascal} or less environment.
This was necessary, as measurements with UCNs have to be done under vacuum to prevent scattering of neutrons by compounds in the air.
Therefore, we developed a holder that allows maintaining the volume around the emulsion at atmospheric pressure.
Finally, a test exposure to UCNs was conducted at Institut Laue-Langevin to obtain the spatial distribution. 

\section{Improvements of the detector}
\label{sec:detector}
\begin{figure}[htbp]
\begin{center}
\includegraphics[scale=0.35,clip]{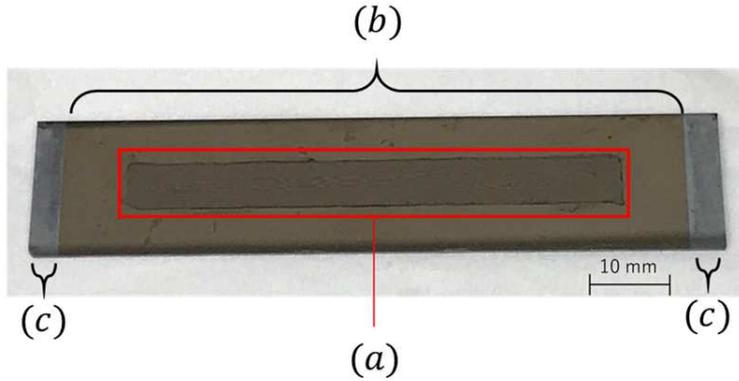}
\caption{Detector used in the measurement of the spatial distribution of quantized UCNs.
($a$) -- ($c$) indicate the following regions: 
($a$) nuclear emulsion,  ($b$) Si substrate and $^{10}$B$_{4}$C-NiC-C layer, ($c$) plain Si substrate.}
\label{fig:detector}
\end{center}
\end{figure}

\begin{figure}[htbp]
\begin{center}
\includegraphics[scale=0.35,clip]{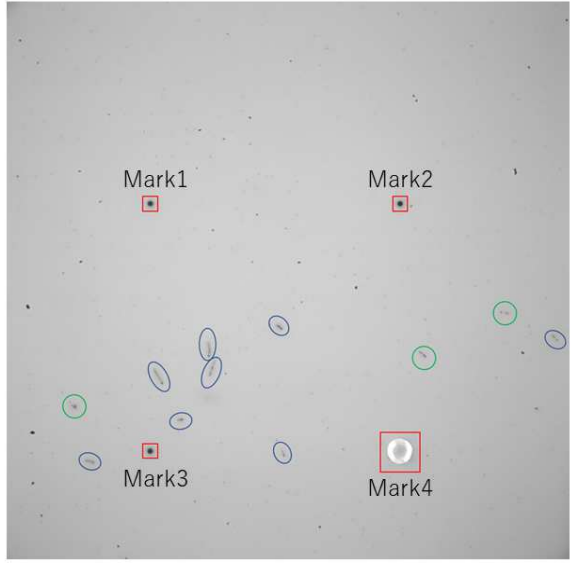}
\caption{Tracks and reference marks on an optical microscope image.
Reference marks have been highlighted by red squares.
There are two different types of markings.
Mark4 in the figure are \SI{5}{\micro \metre} in diameter and lie on a grid with \SI{500}{\micro \metre} spacing.
All other marks (Mark1, Mark2, Mark3) in the figure are \SI{1}{\micro \metre} in diameter and lie on a grid with \SI{50}{\micro \metre} spacing.
$^{7}$Li nuclei and $\alpha$-particles tracks are indicated by blue ellipses.
Green circles indicate stains or tracks at large angles $\theta > \SI{0.9}{\radian}$, which are excluded from the analysis (see main text).}
\label{fig:marks_tracks}
\end{center}
\end{figure}

In order to perform an actual measurement of the spatial distribution of UCNs, we assembled a new emulsion detector (see Figure~\ref{fig:detector}) implementing all of the improvements mentioned above.
We created circular reference marks of \SI{1}{\micro \metre} and \SI{5}{\micro \metre} diameter with an interval of $d_1=\SI{50}{\micro \metre}$ and $d_2=\SI{500}{\micro \metre}$, respectively, over an area of \SI{65x0.2}{\milli\metre} on the Si substrate using an electron beam lithography system (JBX-6300FS).
Absolute coordinate values of the centers of the marks were determined with an accuracy better than \SI{30}{\nano \metre} thanks to laser interferometers used to determine the position during the marking process.
The depth of the marks was \SI{500}{\nano \metre}.
The marks were successfully recognized as black circles under the microscope as shown in Figure~\ref{fig:marks_tracks}.
These marks make it possible to define the entire analysis area in a unified coordinate system.

\begin{figure}[htbp]
\begin{center}
\includegraphics[scale=0.41,clip]{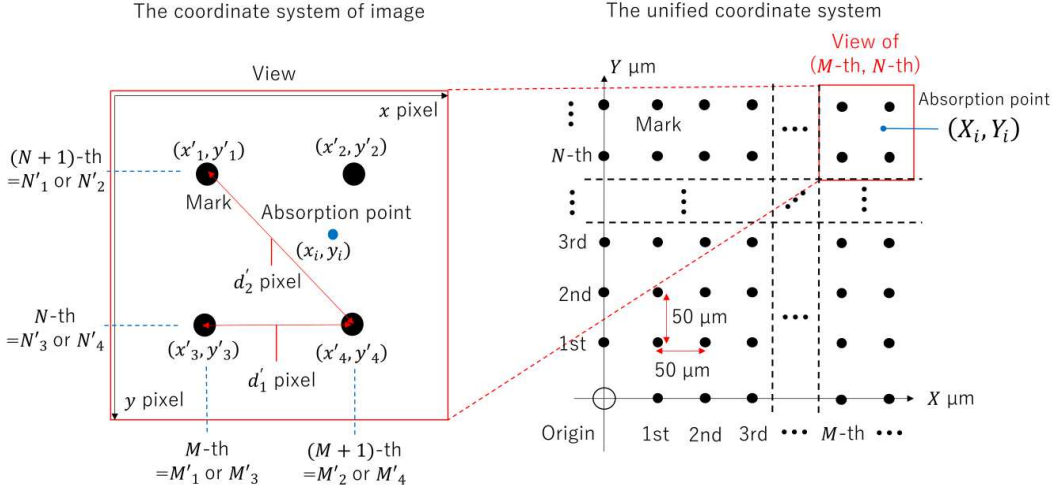}
\caption{Schematic view of the image and unified coordinate systems.
The black circles indicate reference marks.
The blue dot represents an absorption point in both coordinate systems, whose values were $(x_{i}\,\si{pixels},\,y_{i}\,\si{pixels})$ and $(X_{i}\,\si{\micro \metre},\,Y_{i}\,\si{\micro \metre})$.
The coordinates of the marks are $(x'\,\si{pixel},\,y'\,\si{pixel})$, and the subscripts 1, 2, 3, and 4 refer to the upper left, upper right, lower left, and lower right of each view, respectively.
$M$ and $N$ are the lattice indices in $x$ and $y$ directions, respectively, starting from the mark on the origin of the unified coordinate system.
$M'_{1}$ or $M'_{3}$ corresponds to $M$.
$M'_{2}$ or $M'_{4}$ corresponds to $M$+1.
$N'_{3}$ or $N'_{4}$ corresponds to $N$.
$N'_{1}$ or $N'_{2}$ corresponds to $N$+1.}
\label{fig:affine}
\end{center}
\end{figure}

The coordinate values of neutron absorption points were converted from the coordinate system of each image to the unified coordinate system defined by the marks using an affine transformation as follows,
\begin{equation}
\begin{pmatrix}
    X_{i}\\Y_{i}\\1\\
\end{pmatrix} =
\begin{pmatrix}
    A & B & C\\D & E & F\\0 & 0 & 1\\
\end{pmatrix}
\begin{pmatrix}
    x_{i}\\y_{i}\\1\\
\end{pmatrix},
\label{eq:affine}
\end{equation}
where the values of the $i$-th absorption point in the coordinate system of the image are ($x_{i}$ \si{pixel}, $y_{i}$ \si{pixel}), those in the unified coordinate system are ($X_{i}$ \si{\micro \metre}, $Y_{i}$ \si{\micro \metre}) as shown in Figure~\ref{fig:affine}, and $A$, $B$, $C$, $D$, $E$, and $F$ are affine transformation parameters.
The parameters were acquired by the following formula using the information of three marks among four in each view:
\begin{equation}
\begin{pmatrix}
    A_{j}\\B_{j}\\C_{j}\\
\end{pmatrix} =
\begin{pmatrix}
    x'_{j} & y'_{j} & 1\\x'_{k} & y'_{k} & 1\\ x'_{l}& y'_{l} & 1\\
\end{pmatrix}^{-1}
\begin{pmatrix}
    d_{1} M'_{j}\\d_{1} M'_{k}\\d_{1} M'_{l}\\
\end{pmatrix},
\label{eq:affine2}
\end{equation}
\begin{equation}
\begin{pmatrix}
    D_{j}\\E_{j}\\F_{j}\\
\end{pmatrix} =
\begin{pmatrix}
    x'_{j} & y'_{j} & 1\\x'_{k} & y'_{k} & 1\\ x'_{l}& y'_{l} & 1\\
\end{pmatrix}^{-1}
\begin{pmatrix}
    d_{1} N'_{j}\\d_{1} N'_{k}\\d_{1} N'_{l}\\
\end{pmatrix},
\label{eq:affine3}
\end{equation}
where ($j$, $k$, $l$) are (1, 2, 3), (2, 3, 4), (3, 4, 1), or (4, 1, 2), and ($x'_{j}$, $y'_{j}$) are the center coordinates of the ($M'_{j}$, $N'_{j}$)-th mark in ($x$, $y$) directions, respectively, with respect to the origin of the unified coordinate system shown in Figure~\ref{fig:affine}.
$A$, $B$, $C$, $D$, $E$, and $F$ are determined by the average of the $A_{j}$, $B_{j}$, $C_{j}$, $D_{j}$, $E_{j}$, and $F_{j}$ obtained for all ($j$, $k$, $l$) pairs.
We obtained the coordinate values of the marks in two steps.
First, the edge of the mark was acquired using a Canny edge detector~\cite{canny}.
Then, the coordinates of the marks were obtained using a Hough transform~\cite{hough} of the detected edge assuming that marks were circular.
To avoid errors caused by distortion due to the microscope lens, the region inside the marks (\SI{50x50}{\micro \metre}) was set to the center of each tomographic image and the region was called subsequently `view'.
Tomographic images were acquired with an interval of \SI{50}{\micro \metre} in the $x$-$y$ plane.
If deformed markers are created, the respective views are excluded from the analysis since the center coordinates of marks cannot be determined by a Hough transform. 

The error of the transformation from the image coordinate system to the unified coordinate system depends on the error of the center coordinates of the marks.
The error was estimated by measuring the distance between adjacent marks $d'_{1}$ \si{pixel} and those between diagonally adjacent marks $d'_{2}$ \si{pixel} were calculated from $(x'\,\si{pixel},\,y'\,\si{pixel})$ (see Figure~\ref{fig:affine}).
Then, $d'_{1}\,\si{pixel}$ and $d'_{2}\,\si{pixel}$ were translated to $d_{1}\,\si{\micro \metre}$ and $d_{2}\,\si{\micro \metre}$, respectively, by using the fact that one pixel is equal to \SI{55}{\nano \metre}.
$d{_1}$ was measured for 400 marks and $d{_2}$ for 200 marks, resulting in average values $d_{1} =  \SI{50.0010 \pm 0.0019}{\micro \metre}$ and $d_{2} = \SI{70.7046 \pm 0.0029}{\micro \metre}$ and standard deviations $\sigma_{d_{1}} =  \SI{38.3 \pm 1.4}{\nano \metre}$ and $\sigma_{d_{2}} = \SI{40.7 \pm 2.0}{\nano \metre}$, obtained by fitting a Gaussian function to the histograms, respectively.
Assuming the errors of the single mark coordinates to be equal, we divide the errors in $d_1\,\si{\micro\metre}$ and $d_2\,\si{\micro\metre}$, by $\sqrt{2}$ to estimate the errors of the mark center coordinates.
Conservatively, using the larger value $\sigma_{d_{2}}$, we thus obtain the standard deviation of the center coordinates of the marks to be \SI{28.8 \pm 1.4}{\nano \metre}.
Note that this standard deviation includes not only the limited accuracy of determining the position of the marks during the electron beam lithography but also the accuracy of reading the coordinate values of the marks.
The given value (\SI{28.8 \pm 1.4}{\nano \metre}) is consistent with the specified accuracy (less than \SI{30}{\nano \metre}) of our electron beam lithography equipment.

\begin{figure}[htbp]
\begin{center}
\includegraphics[scale=0.35,clip]{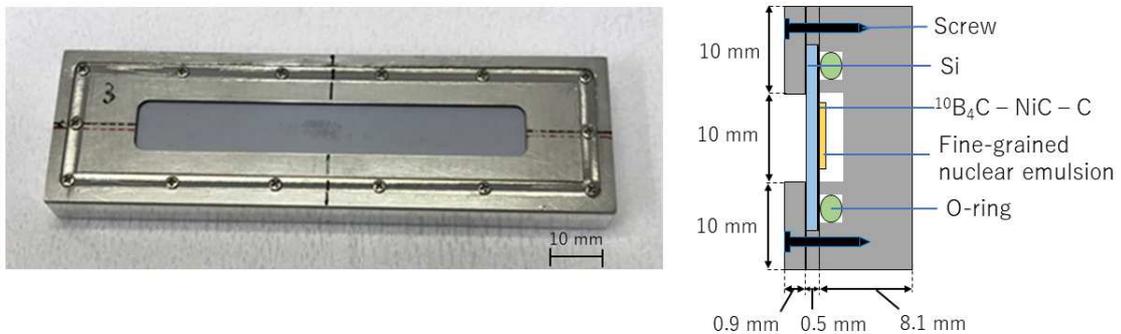}
\caption{Holder allowing in-vacuum use of the emulsion detector.
(left: photograph of the holder, right: schematic view of its cross section).
The central part of the top cover of the holder was hollowed out to allow direct exposure of the surface of Si to vacuum.
The yellow part of the schematic view shows the nuclear emulsion layer; the volume around it is kept at atmospheric pressure.}
\label{fig:holder}
\end{center}
\end{figure}

The second issue solved by the new detector relates to measurements in a vacuum environment.
If the nuclear emulsion would be exposed to a vacuum, the contained water would disperse quickly, causing cracking and peeling of the emulsion surface.
The torsion would further lead to an increase in silver grains not related to tracks that would be recognized as noise (fog) in the image.
To operate the emulsion in a vacuum nonetheless, we developed a holder to maintain the volume around it at atmospheric pressure (Figure~\ref{fig:holder}).
The central part of the top cover was hollowed out in an area of \SI{10x80}{\milli \metre} to prevent scattering of incident ultracold neutrons at the holder.
The nuclear emulsion gel was applied to the central area of \SI{7x65}{\milli \metre} to prevent contact with the holder.
We conducted a durability test by placing a nuclear emulsion in this holder under a vacuum of \SI{3e-3}{\pascal} for two days, corresponding to a typical duration for an exposure with ultracold neutrons.
Subsequently, the state of the nuclear emulsion was confirmed with an optical microscope after development.
Neither cracks nor the increase of fog was observed on the surface of it, which confirms that the holder serves its purpose.

\section{Measurement of the resolution using a gadolinium grating}
\label{sec:Gd_analysis}
\begin{figure}[htbp]
\begin{center}
\includegraphics[scale=0.4,clip]{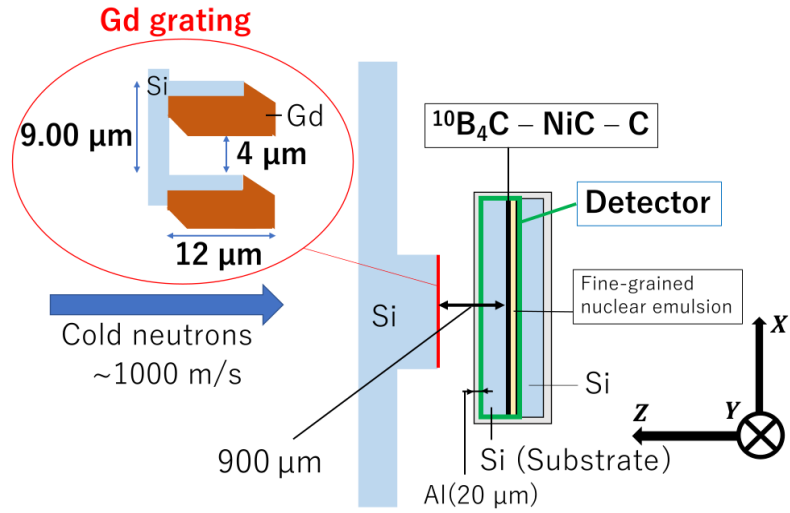}
\caption{Setup for the measurement of cold neutron detection resolution.
The Gd grating is located on the downstream side of a Si-slab placed in front of the detector.
A Si plate was inserted to separate the emulsion layer from the Al foil to avoid the chemical effects of Al on the emulsion.}
\label{fig:setup_Gd}
\end{center}
\end{figure}

\begin{figure}[htbp]
\begin{center}
\includegraphics[scale=0.43,clip]{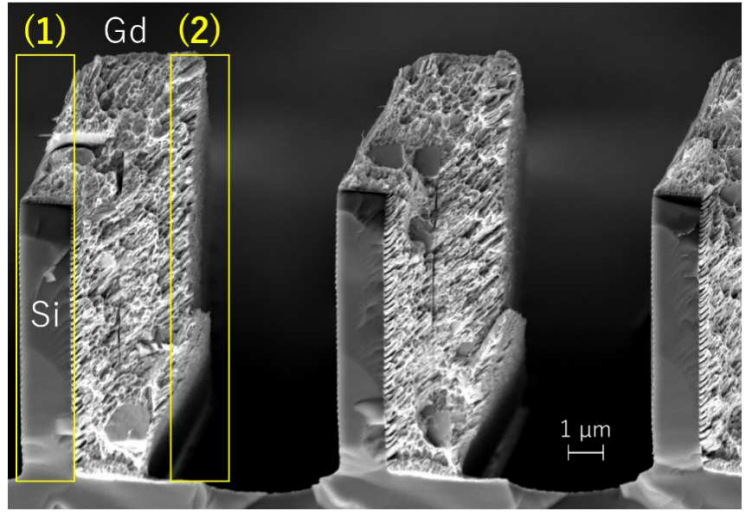}
\caption{Electron micrograph showing a cross section of a Gd grating that was manufactured by the same process as that used in this experiment.
The thickness of Gd in (1) is not sufficient to prevent cold neutrons from passing.
We used the transmission image of the edge (2) for the estimation of the resolution of the detector.}
\label{fig:Gd_grating}
\end{center}
\end{figure}

The spatial resolution of the improved detector was evaluated by acquiring a transmission image of a neutron-absorbing Gd grating that has a periodic structure along the $X$-axis direction (see Figure~\ref{fig:setup_Gd}) and creates a sharp shadow on the detector.
Subsequently, the edges of the shadow were numerically analyzed by fitting the profile with an error function.
This experiment was conducted at BL05~\cite{BL5_1,BL5_2} of the Materials and Life Science Experimental Facility in J-PARC using cold neutrons of approximately \SI[per-mode=symbol]{1000}{\metre\per\second}.
The \SI{12}{\micro \metre} thick grating was created by vapor deposition of Gd from an oblique angle on a Si mold with a comb-like structure of period of \SI{9.00}{\micro \metre} (measured by confocal microscopy), and the opening of the Gd grating was \SI{4}{\micro \metre}~\cite{Gd_grating}.
Figure~\ref{fig:Gd_grating} shows an electron micrograph of the cross section of a similar Gd grating made by the same process.
A nuclear emulsion wrapped with two aluminum foils with a total thickness of \SI{20}{\micro \metre} for light shielding was placed behind the Gd grating.
The detector was then exposed through the Gd grating for \SI{2.0e4}{\second} using cold neutrons (Figure~\ref{fig:setup_Gd}) with a beam flux of \SI[per-mode=symbol]{7.6e4}{neutrons\per\centi \metre\squared\per\second}, as measured with a $^{3}$He proportional counter.
We estimated the transmittance of the \SI{12}{\micro \metre} Gd layer to be \SI{2}{\percent}, which can be ignored. 
Due to the geometric view factor given by the Gd grating, one-third of the total beam reached the detector.
From the total number of impeding neutrons, and the track count on the detector we determined a detection efficiency of \SI{0.50}{\percent}.
This corresponds well to the theoretical estimate \SI{0.65}{\percent} (corresponding to \num{3.3e2} tracks per \SI{100x100}{\micro \metre}) based on the thickness of the $^{10}$B$_{4}$C layer.
The beam divergence angle was \SI{5.6e-2}{\milli \radian} and \SI{0.56}{\milli \radian} in the $X$-axis and $Y$-axis direction, respectively.

\begin{figure}
\begin{center}
\includegraphics[scale=0.35,clip]{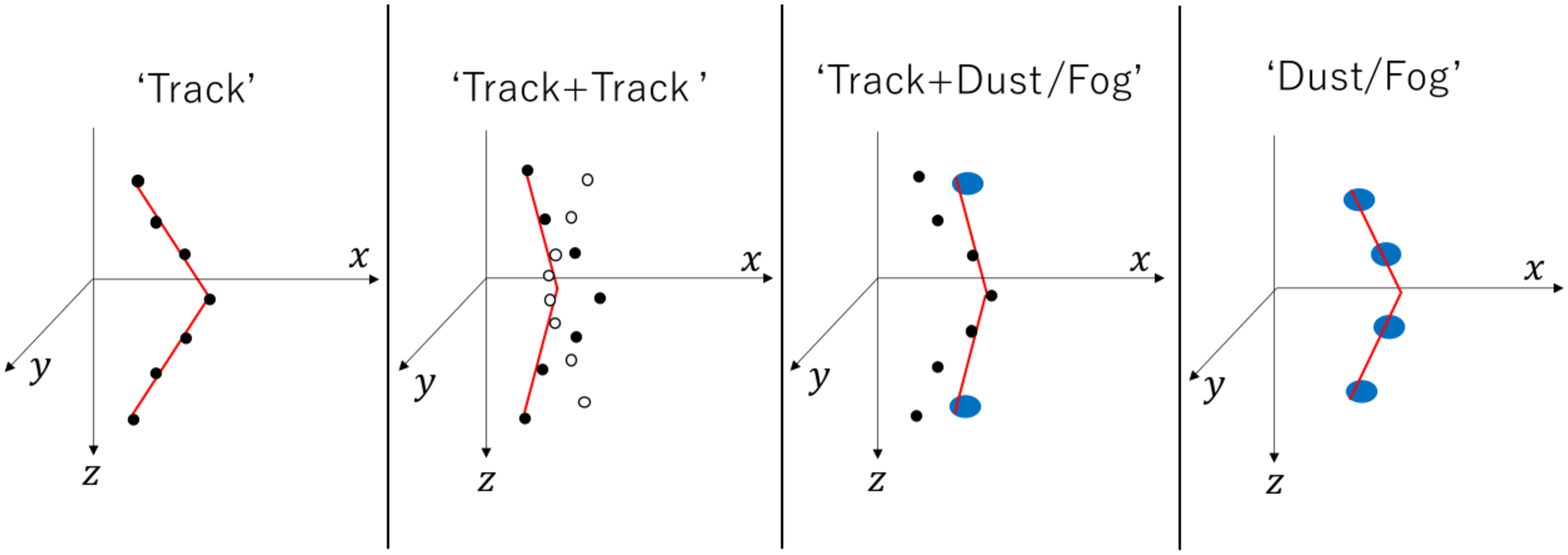}
\caption{Classification of the reconstructed tracks.
The reconstructed tracks were classified into four types.
Red lines show reconstructed tracks.
Black dots show silver particles.
White dots of `Track+Track' show silver particles on a track originally belonging to a different track than that of black dots.
Blue ellipses of `Track+Dust/Fog' or `Dust/Fog' show fog or dust.}
\label{fig:classification}
\end{center}
\end{figure}

\begin{figure}[htbp]
\begin{center}
\includegraphics[scale=0.33,clip]{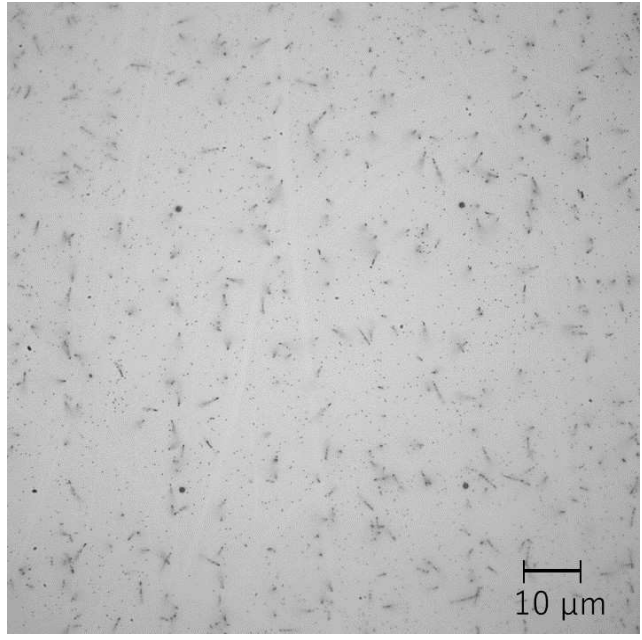}
\caption{Transmission image of Gd grating.}
\label{fig:Gd_trans_image}
\end{center}
\end{figure}

Figure~\ref{fig:Gd_trans_image} shows an image of the detector taken with an optical microscope after development, in which the striped distribution of tracks can clearly be seen.
We acquired the absorption points using the analysis method explained in Section~\ref{sec:intro}.
The reconstructed tracks were classified into four types: `Track', `Track+Track', `Track+Dust/Fog', and `Dust/Fog' as shown in Figure~\ref{fig:classification}.
`Track' means that all grains belong to an original track.
`Track+Track' is an erroneously reconstructed track that consists of grains originally belonging to different tracks.
`Track+Dust/Fog' means that some grains of the track are fog or dust origin.
`Dust/Fog' means that all grains are dust or fog origin.
The types other than `Track' were treated as noise.
Since the grains of a track that was categorized as noise are not distributed on a straight line, it is considered that the $\chi^{2}/\mathrm{ndf}$ of the corresponding line fit is increased with respect to a regular track. 
Therefore, the noise contamination rate can be reduced by not using reconstructed tracks with a large value of $\chi^{2}/\mathrm{ndf}$ in the analysis.
$\chi^{2}$ was acquired by the following formula,
\begin{equation}
\chi^{2} = \sum_i\left(\left(\frac{\delta R_i}{\delta a\sqrt{1+\left(\frac{\delta b}{\delta a}\right)^2\tan^2{\theta}}}\right)^2+\left(\frac{\delta L_i}{\delta a}\right)^2\right)
\label{eq:chi2},
\end{equation}
where
\begin{equation}
\begin{pmatrix}
\delta R_{i} \\ \delta L_{i}
\end{pmatrix}
 = 
\begin{pmatrix}
\cos{\phi} && \sin{\phi} \\ -\sin{\phi} && \cos{\phi}
\end{pmatrix}
\begin{pmatrix}
x_{t}(z_{i})-x_{i} \\ y_{t}(z_{i})-y_{i}
\end{pmatrix}
\label{eq:lat_rad}.
\end{equation}
Here, $\theta$ and $\phi$ are the angle of a track in a spherical coordinate system, the coordinates of the $i$-th grain are ($x_{i}$, $y_{i}$, $z_{i}$), those of the reconstructed track at $z_{i}$ are ($x_{t}(z_{i})$, $y_{t}(z_{i})$), $\delta a=\SI{57}{\nano \metre}$ is the standard deviation in the $\delta L_{i}$ direction , and $\delta b=\SI{300}{\nano \metre}$ is the standard deviation in the $z$ direction.
In order to systematically remove noise from our data, we introduce the empirical criterion $\chi^{2}/\mathrm{ndf} < 2$.
This cutoff value was determined by analyzing the typical $\chi^{2}/\mathrm{ndf}$ for reconstructed tracks belonging to the four categories described above (see also Figure~\ref{fig:classification}).
For this purpose, we randomly selected 20 reconstructed tracks and classified them according to their respective $\chi^{2}/\mathrm{ndf}$ as shown in Table~\ref{tab:clas_track}.
The number of survived tracks after applying the cutoff was 3022.

\begin{table}[htbp]
\centering
\caption{Classification of 20 randomly selected reconstructed tracks according to their $\chi^{2}/\mathrm{ndf}$, used to empirically set the threshold for acceptance $\chi^2/\mathrm{ndf} < 2$.
These tracks were visually checked and classified as described in the main text.\\
}
  \begin{tabular}{l|cccc}
    $\chi^{2}/\mathrm{ndf}$ & 0.00 -- 0.67 & 0.67 -- 1.33 & 1.33 -- 2.00 & 2.00 -- 2.67\\ \hline
    Total events & 939 & 1267 & 816  & 529\\ \hline  \hline
    Track & 17 & 16 & 14 & 5\\
    Track+Track & 1 & 2 & 3 & 10\\
    Track+Dust/Fog & 2 & 0 & 2 & 5\\
    Dust/Fog & 0 & 2 & 1 & 0\\ \hline
    Checked tracks & 20 & 20 & 20 & 20 
  \end{tabular}
  \label{tab:clas_track}
\end{table}

\begin{figure}[htbp]
\begin{center}
\includegraphics[scale=0.45,clip]{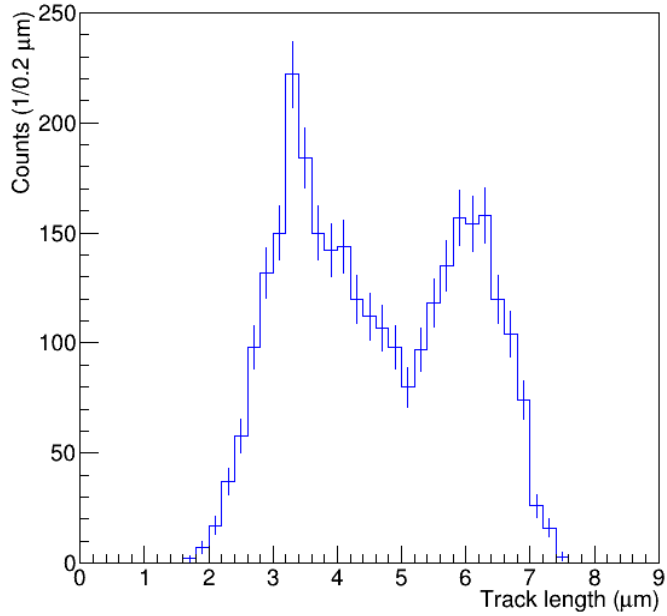}
\caption{Distribution of the length of survived tracks in the analysis.
The figure shows the length from the absorption points to the ending points of the grain, assuming the absorption reaction happens in the middle of the thickness of $^{10}$B$_{4}$C layer.
The two peaks of the distribution correspond to ranges of tracks of $^{7}$Li nuclei and $\alpha$-particles with theoretical center values of \SI{2.7 \pm 0.4}{\micro \metre} and \SI{5.2 \pm 0.4}{\micro \metre} in the emulsion, calculated by SRIM-2013, respectively.
Slight differences from the calculated values are explained with the error of the shrinkage rate of the emulsion layer during the development.
}
\label{fig:track_length}
\end{center}
\end{figure}

\begin{figure}[htbp]
\begin{center}
\includegraphics[scale=0.42,clip]{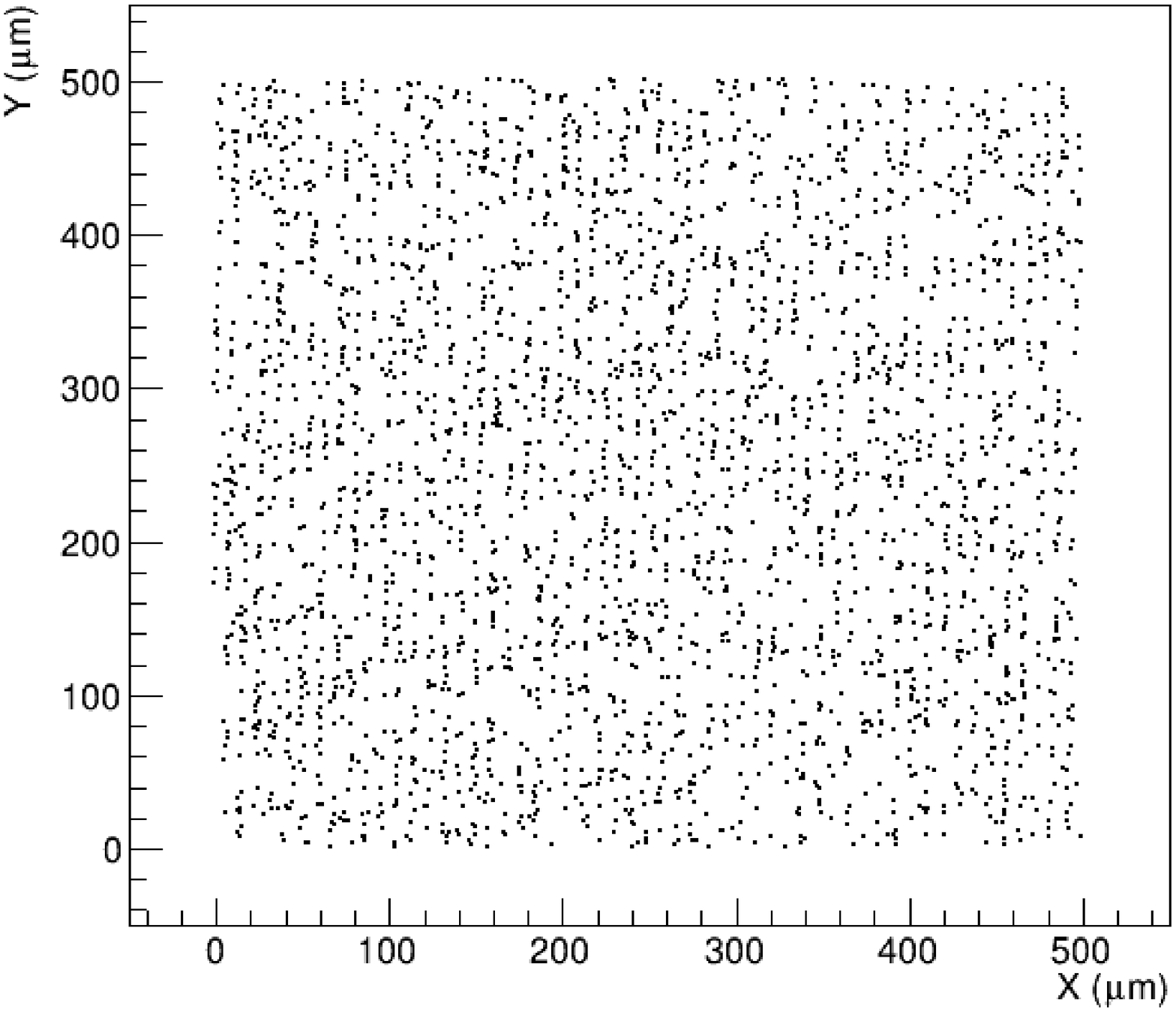}
\caption{Distribution of absorption points of neutrons that passed through the Gd grating.}
\label{fig:Gd_trans_2d}
\end{center}
\end{figure}

The distribution of the length of survived tracks used in the analysis is shown in Figure~\ref{fig:track_length}.
The analyzed region was \SI{500x500}{\micro \metre}, and the coordinates of the absorption points obtained in each view was converted into the unified coordinate system using Equation~\ref{eq:affine}.
Figure~\ref{fig:Gd_trans_2d} shows a plot of the absorption points.
A striped pattern corresponding to that of the structure of the Gd grating can be observed in the plot, as expected.
In order to remove an overall tilt angle $\alpha$ and determine the period $d$ of the pattern, we use the Rayleigh test function $R(\alpha, d)$~\cite{rayleigh_1,rayleigh_2}:
\begin{equation}
R(\alpha, d) = \frac{2}{N}\left(\left(\sum_{i}\sin\left(\frac{2\pi x_i(\alpha)}{d}\right)\right)^{2}+\left(\sum_{i}\cos\left(\frac{2\pi x_i(\alpha)}{d}\right)\right)^{2}\right)\,,
\label{eq:Rayleigh_1}
\end{equation}
where
\begin{equation}
x_{i}=X'_{i}\cos\alpha+Y'_{i}\sin\alpha+x_{m}\,,
\label{eq:Rayleigh_2}
\end{equation}
and
\begin{equation}
\begin{pmatrix}
    X'_{i}\\Y'_{i}\\
\end{pmatrix}
 =
\begin{pmatrix}
    X_{i}-x_{m}\\Y_{i}-y_{m}\\
\end{pmatrix}.
\label{eq:Rayleigh_3}
\end{equation}
Here, $N$ is the total number of absorption points, ($X_{i}$, $Y_{i}$) are the coordinates in the unified coordinate system [c.f. Equation~\ref{eq:affine}], and ($x_{m}$, $y_{m}$) are the average values of the coordinates of all absorption points.
Optimum values $\alpha_{m}$ = \SI{-1.623 \pm 0.030e-2}{\radian} and $d_{m}$ =  \SI{8.9997 \pm 0.0026}{\micro \metre} were obtained by maximizing $R(\alpha = \alpha_{m}, d = d_{m})$ numerically.
The detected period is consistent with the designed value of the Gd grating, \SI{9.00}{\micro \metre}.

\begin{figure}[htbp]
\begin{center}
\includegraphics[scale=0.35,clip]{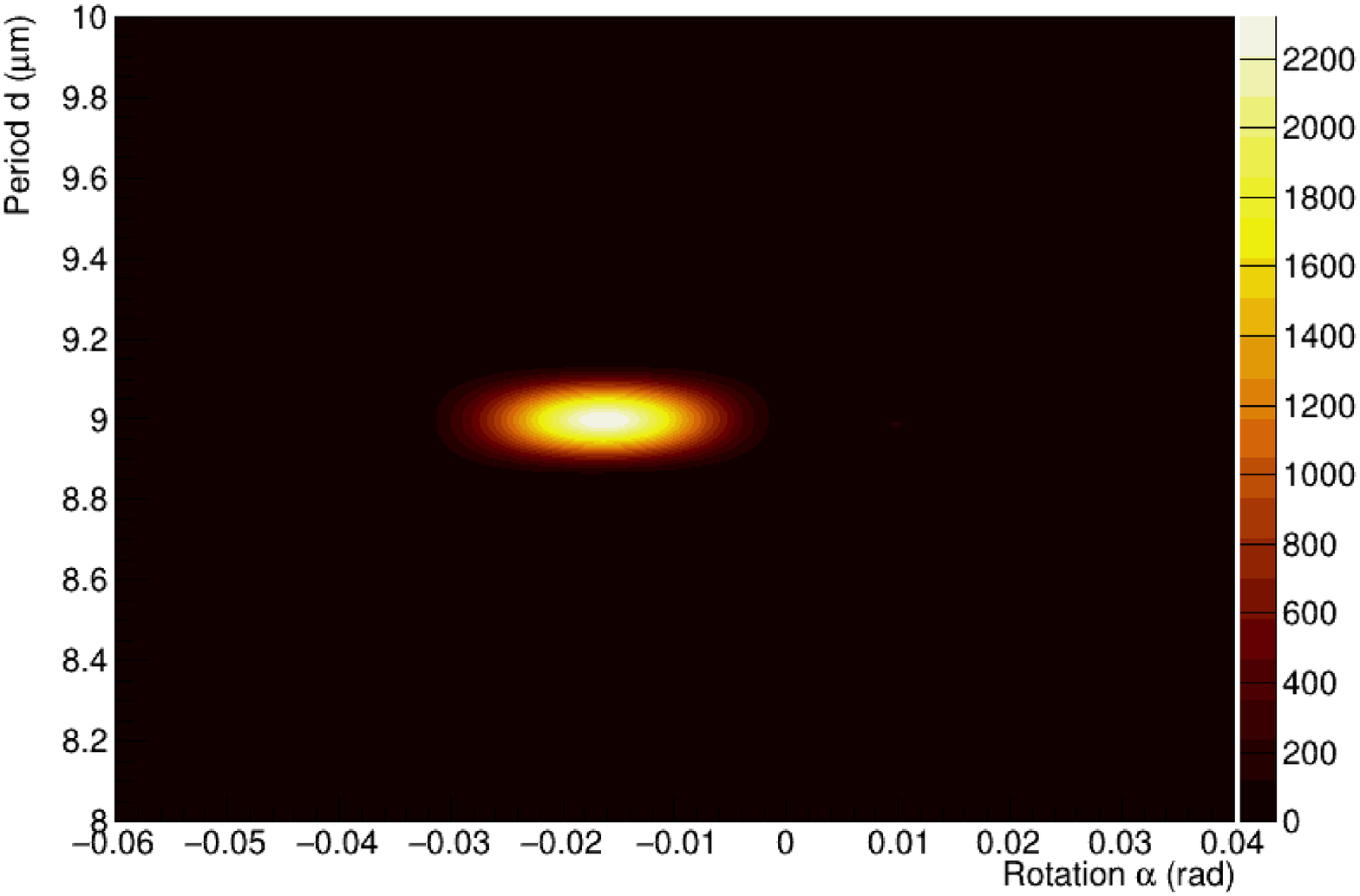}
\caption{
Results of the Rayleigh test.
The color coding expresses the value of the Rayleigh test function $R(\alpha, d)$.
The brightest $R$ point indicates the optimum values of $\alpha$ and $d$.}
\label{fig:Rayleigh}
\end{center}
\end{figure}

\begin{figure}[htbp]
\begin{center}
\includegraphics[scale=0.4,clip]{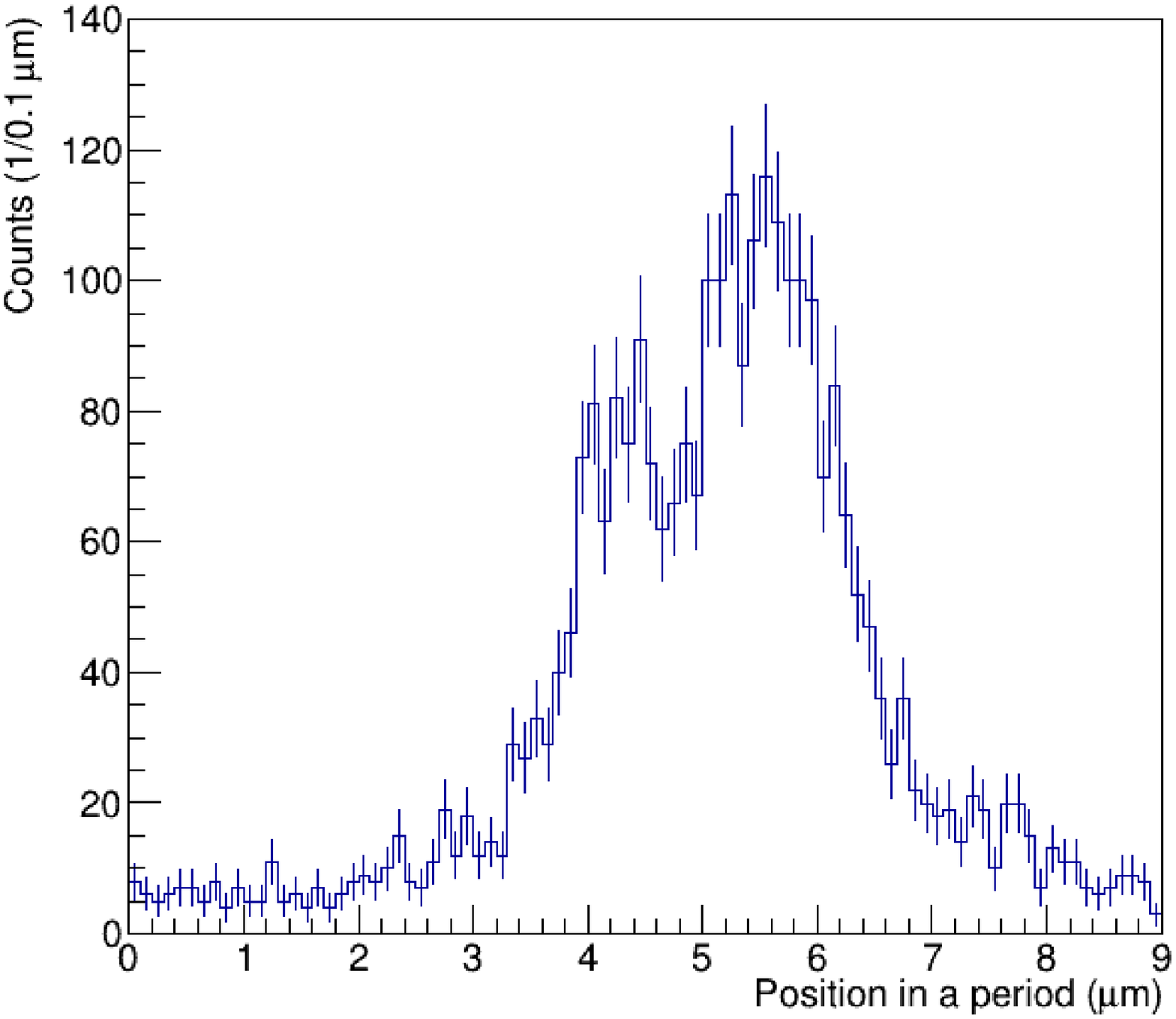}
\caption{Distribution of $x_i$ modulo $d_{m}$ in the Gd transmission measurement.
The resolution of the detector was evaluated using $x_i \text{ modulo } d_m > \SI{5.0}{\micro \metre}$ whose region corresponded to the sharp edge of Gd grating.}
\label{fig:hist_Gd_trans}
\end{center}
\end{figure}

\begin{figure}[htbp]
\begin{center}
\includegraphics[scale=0.4,clip]{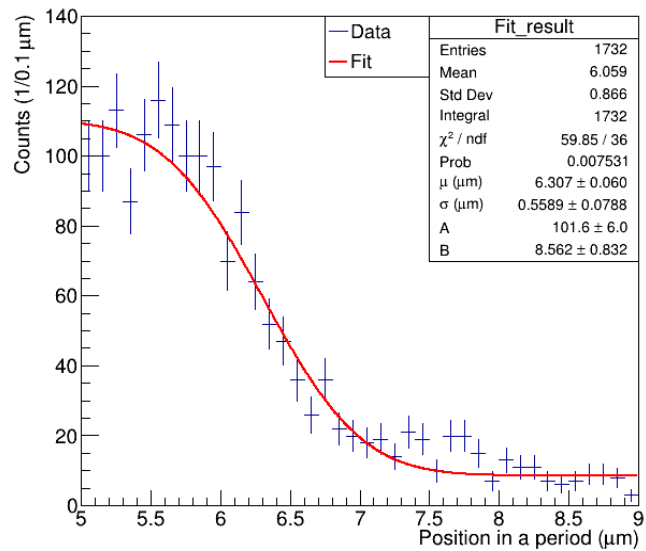}
\caption{Fit result of the absorption points around the Gd grating edge.
The blue histogram data is an excerpt of the data in Figure~\ref{fig:hist_Gd_trans} for $x_i \text{ modulo } d_m > \SI{5.0}{\micro \metre}$.
The red line shows the result of the least squares fit of  Equation~\ref{eq:fit_Gd} to the data. }
\label{fig:results_resolution}
\end{center}
\end{figure}

Subsequently, the tilt was removed from the data by rotating coordinates by $-\alpha_{m}$ using Equation~\ref{eq:Rayleigh_2}.
The histogram of the distribution of the absorption points folded with the period of $d_{m}$, i.e., $x_{i}$ modulo $d_{m}$, is given in Figure~\ref{fig:hist_Gd_trans}.
A peak corresponding to the opening of the Gd grating was observed from 3.5 to \SI{7.0}{\micro \metre}.
In addition, it was understood that the part from 4 to \SI{5}{\micro \metre} corresponded to (1) in Figure~\ref{fig:Gd_grating} because the number of events from 4 to \SI{5}{\micro \metre} of the horizontal axis in Figure~\ref{fig:hist_Gd_trans} was smaller than the number of events from 5 to \SI{6}{\micro \metre}.
The transmission image corresponding to (1) is more blurred than that corresponding to (2) since the thickness of Gd in (1) is not sufficient.
Therefore, we did not use the left side of the transmission image but the right side of it.
We fitted the right edge of the data in Figure~\ref{fig:hist_Gd_trans} with the error function:
\begin{equation}
f(x)=A\left(\frac{1}{\sqrt{2\pi}\sigma}\int_x^\infty\exp{\left(-\frac{(t-\mu)^2}{2\sigma^2}\right)}\,\mathrm{d}t\right)+B,
\label{eq:fit_Gd}
\end{equation}
where $\mu$ \si{\micro \metre} is the coordinate of the edge of a stripe on the detector, $\sigma$ \si{\micro \metre} is the effective blurring of the distribution, $A$ is a scaling factor of the distribution, and $B$ is the amount of background noise.
All of the mentioned variables were used as free parameters of the fit.
Since the cross section of the sputtered Gd layer is not a perfect rectangle as shown in Figure~\ref{fig:Gd_grating}, it is assumed that $\sigma$ becomes larger than a certain value.
Assuming that all Gd shapes are the same as in Figure~\ref{fig:Gd_grating}, the thickness distribution can be obtained from the figure.
The lower limit of $\sigma$ was \SI{0.14}{\micro \metre} by fitting the transmission image estimated from the thickness distribution with Equation~\ref{eq:fit_Gd}.
The fitting result is shown in Figure~\ref{fig:results_resolution}.
We obtain a spread of $\sigma=\SI{0.56 \pm 0.08}{\micro \metre}$, which includes the effects of the intrinsic resolution of the detector, the accuracy of the connecting of views using the reference marks, the spread of the beam, the shape of the Gd grating, and the variation in the amount of Gd on each Si mold.
It is therefore reasonable to state that the intrinsic resolution of the detector is expected to be better than the given value.
The divergence angle of the beam in the $x$-axis direction (horizontal axis in Figure~\ref{fig:hist_Gd_trans}) was obtained by rotating the original divergence angle by $-\alpha_{m}$.
The maximum value was \SI{6.5e-2}{\milli \radian}.
The distance between the Gd grating and the $^{10}$B$_{4}$C layer of the detector was \SI{900}{\micro \metre}; hence, the spread of the transmission image due to the divergence angle was \SI{60}{\nano \metre}.
It was considered that probably the variation in the amount of Gd on each Si mold was dominant for the value of the resolution. 

\newpage
\section{Measurement of the spatial distribution of quantized UCNs}
\label{sec:UCNs_analysis}
\begin{figure}[htbp]
\begin{center}
\includegraphics[scale=0.3,clip]{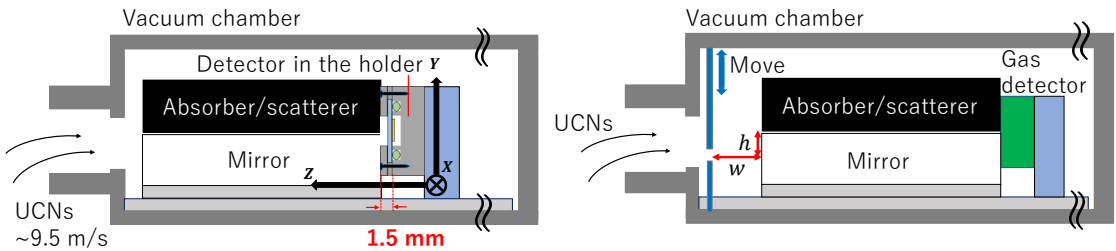}
\caption{Setup for obtaining the quantized spatial distribution of UCNs.
The figures on the left and right show the schematic view of the setup of the measurement of coordinates and velocity of UCNs, respectively.
The distance from the exit of the mirror and absorber/scatterer to $^{10}$B$_{4}$C layer is \SI{1.5}{\micro \metre}.}
\label{fig:setup_ILL}
\end{center}
\end{figure}

\begin{figure}[htbp]
\begin{center}
\includegraphics[scale=0.4,clip]{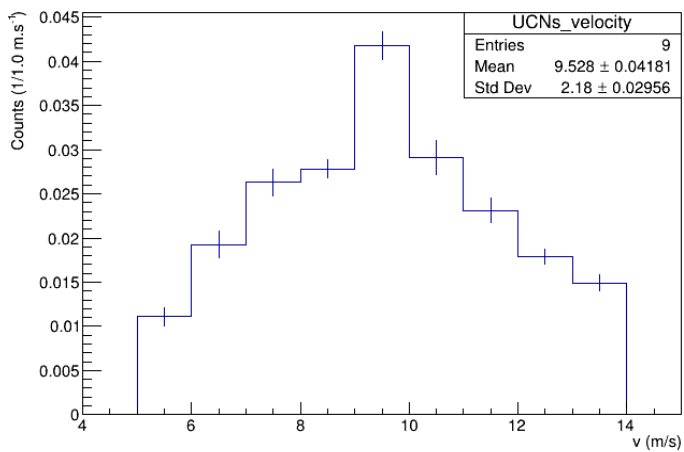}
\caption{Measured velocity distribution at the UCN beam of the PF2 installation.
The counts of UCNs of 4 -- \SI[per-mode=symbol]{11}{\metre\per\second} were measured using a gas detector with a $^{10}$B layer.}
\label{fig:velocity}
\end{center}
\end{figure}

As a demonstration of our emulsion detector, we attempted to measure the spatial distribution of UCNs that take quantized states in the potential well created between the Pseudo-Fermi potential of a horizontal flat mirror and the gravitational potential of the Earth, a system commonly known as `quantum bouncer'~\cite{10.1119/1.10024}.
This experiment was conducted at PF2 of the Institut Laue-Langevin.
The experimental setup is shown in Figure~\ref{fig:setup_ILL}.
A combination of a flat mirror with a rough absorber/scatterer on top, separated by precision spacers of nominal thickness \SI{30}{\micro\metre} creates a selector for vertical energy states of neutrons~\cite{10.1016/j.nima.2009.07.073}.
At the exit of the selector, neutrons arrive in one of the lowest few gravitational quantum states~\cite{jenke2013_10B_CR39}.
The mirror, absorber/scatterer and vacuum chamber in Figure~\ref{fig:setup_ILL} were taken from the \textit{q}\textsc{Bounce} experiment~\cite{short_for_exp3}.
Both the flat mirror and the rough absorber/scatterer are made of borosilicate glass.
The size of the mirror was \SI{15}{\centi \metre} and \SI{20}{\centi \metre} in the $Z$- and $X$-axes, respectively.
The size of the absorber/scatterer was \SI{15}{\centi \metre} and \SI{10}{\centi \metre} in the $Z$-axis and the $X$-axis directions, respectively.
All parts of the setup were installed in a vacuum chamber, and the detector was exposed with UCNs for 16 hours in a vacuum of \SI{1}{\pascal} or less, where the scattering by air can be neglected.
The beam flux was estimated to be \SI[per-mode=symbol]{0.21}{neutrons\per\second} for the region exposed to neutrons from an actual measurement using a gas detector with a $^{10}$B layer~\cite{jenke2013_10B_CR39}.
The velocity distribution of the UCNs was measured using an aperture system as shown in Figure~\ref{fig:velocity}.
Two plates of borated Al create a slit at horizontal and vertical distances $w$ and $h$, respectively.
The slit together with the gap between the absorber/scatterer and flat mirror define a flight parabola that selects the horizontal velocity $v$ according to 
\begin{equation}
v = \sqrt{\frac{gw^2}{2h}}\,,
\label{eq:velocity}
\end{equation}
where $g$ is the gravitational acceleration.
The velocity distribution was measured by varying $h$ and counting the admitted UCNs by a gas detector as shown in the left of Figure~\ref{fig:velocity}.
The result of this measurement is plotted on the right of Figure~\ref{fig:velocity} and shows an almost Gaussian distribution with a mean of \SI[per-mode=symbol]{9.53}{\metre\per\second} and a standard deviation of \SI[per-mode=symbol]{2.18}{\metre\per\second}.

We reconstructed the tracks and classified them as explained in Section~\ref{sec:Gd_analysis}.
The result of the classification and the number of total events are summarized in Table~\ref{tab:clas_track_2}.
The reason for the noise contamination rate here being lower than in the analysis of Gd grating is that the tracks are less likely to overlap in this measurement due to the smaller density of tracks on the detector.

\begin{figure}[htbp]
\begin{center}
\includegraphics[scale=0.4,clip]{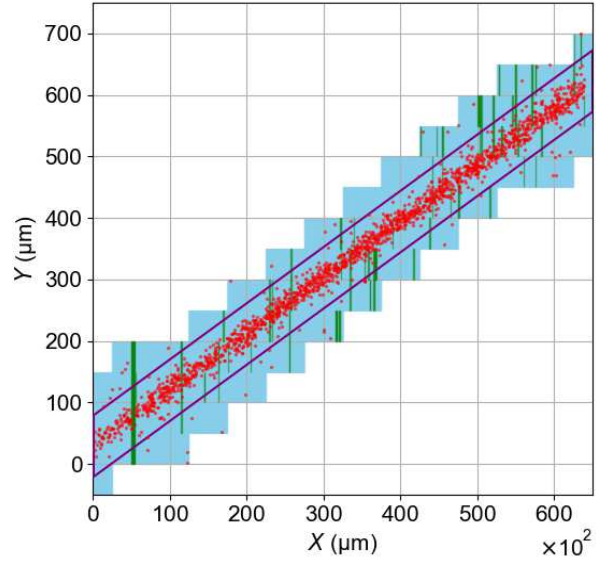}
\caption{Scanned area and the distribution of absorption points.
The length of the vertical axis is 1/100 of the length of the horizontal axis.}
\label{fig:scanned_area}
\end{center}
\end{figure}

Figure~\ref{fig:scanned_area} shows the distribution of the scanned area and the absorption points (red) of neutrons.
The area shown in light blue in Figure~\ref{fig:scanned_area} shows the scanned region; green indicates areas that were excluded due to difficulties in the recognition of marks due to deformed reference marks or the presence of large dust particles on the surface of the C layer.
The scanned area was \SI{65}{\milli \metre} and \SI{0.2}{\milli \metre} in the directions of the $X$- and $Y$-axes, respectively.
The excluded area was \SI{2.7}{\percent} of the scanned area.
We restricted our analysis to absorption points inside the purple lines, containing a total of 1821 tracks.

\begin{table}[htbp]
\centering
\caption{Classification of 20 randomly selected reconstructed tracks depending on their $\chi^{2}/\mathrm{ndf}$.
These tracks were visually checked and classified as described in the main text.\\}
  \begin{tabular}{l|cccc}
    $\chi^{2}/\mathrm{ndf}$ & 0.00 -- 0.67 & 0.67 -- 1.33 & 1.33 -- 2.00 & 2.00 -- 2.67\\ \hline
    Total events & 729 & 830 & 314 & 175 \\ \hline \hline
    Track & 20 & 20 & 19 & 14\\
    Track+Track & 0 & 0 & 0 & 0\\
    Track+Dust/Fog & 0 & 0 & 0 & 1\\
    Dust/Fog & 0 & 0 & 1 & 5\\ \hline
    Checked tracks & 20 & 20 & 20 & 20\\ 
  \end{tabular}
  \label{tab:clas_track_2}
\end{table}

\begin{figure}[htbp]
\begin{center}
\includegraphics[scale=0.45,clip]{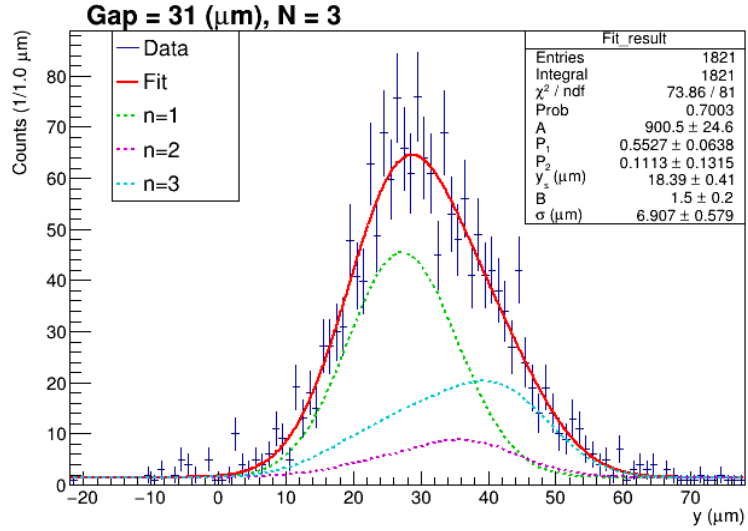}
\caption{Fit result using Equation~\ref{eq:fit_fun_fin_res}, which includes the effect of blurring.
The effective gap was fixed to \SI{31}{\micro \metre} and the maximum number of the quantum state was 3.
The blue histogram shows the distribution of obtained absorption points.
The red line shows the best fit result obtained with model $f(y)$. 
Dashed lines of green, purple, and light blue show the probability distributions of the ground state, the first excited state, and the second excited state, respectively.}
\label{fig:fit_result_combo}
\end{center}
\end{figure}

The obtained line of absorption points was slightly tilted with respect to the horizontal axis as shown in Figure~\ref{fig:scanned_area}, due to the setting accuracy of the detector.
From the least squares fit of the coordinates to a linear function, we obtained a rotation of \SI{9.146 \pm 0.017}{\milli \radian} and a small vertical offset that both were removed from the data.
We created a histogram of the data along the vertical $y$ direction shown in Figure~\ref{fig:fit_result_combo}, where the effect of the excluded area was considered.
We compared the obtained distribution with the theoretical curve to verify the precision of the measurement and estimate the ratio of each quantum state.
Neutrons bound by the pseudo-Fermi potential $V_F$ of a horizontal surface from below and the gravitational potential $mgy$ rising linearly with the distance $y$ above the surface obey the Schr\"odinger equation~\cite{10.1119/1.10024}
\begin{equation}
E\psi(y)=\left(-\frac{\hbar^2}{2m}\frac{\partial^2}{\partial y^2}+mgy + V_F\Theta(y_s-y)\right)\psi(y),
\label{eq:sch_eq}
\end{equation}
where $\psi$ is the wave function, $y_s$ is the vertical coordinate of the mirror surface, $\Theta$ is the Heaviside step function, $m$ is the mass of a neutron, $E$ is the energy of the system, $\hbar$ is the reduced Planck constant, and $g$ is the gravitational acceleration.
Since $V_F$ is several orders of magnitude larger than the neutron's kinetic energy in the vertical direction, we can safely assume the boundary condition
\begin{equation}
\psi(y=y_s)=0
\label{eq:boundary_1},
\end{equation}
for which neutrons cannot enter the mirror.
The same is true for the rough surface of the absorber/scatterer.
However, due to the stochastic roughness of this surface, an effective gap size $d_e$ has to be considered at this boundary,
\begin{equation}
\psi(y=y_s + d_e)=0.
\label{eq:boundary_2}
\end{equation}
When Equation~\ref{eq:sch_eq} is solved under the boundary conditions of Equation~\ref{eq:boundary_1} and~\ref{eq:boundary_2}, the eigenfunction of the $n$-th state can be written as,
\begin{equation}
\psi_n(y)=C_n \mathrm{Ai}\left(\frac{y}{y_0}-\frac{E_n}{E_0}\right)+D_n \mathrm{Bi}\left(\frac{y}{y_0}-\frac{E_n}{E_0}\right),
\label{eq:psi}
\end{equation}
where $y_0$ and $E_0$ are the characteristic scale and energy for the gravitational quantum states, respectively;
\begin{equation}
y_0=\left(\frac{\hbar^2}{2m^2g}\right)^\frac{1}{3} \approx \SI{5.9}{\micro \metre}
\label{eq:boundary_3}
\end{equation}
and
\begin{equation}
 E_{0}=mgy_{0} \approx \SI{0.60}{\pico \electronvolt}.
\label{eq:boundary_4}
\end{equation}
Here, Ai and Bi denote the Airy functions, $C_{n}$ and $D_{n}$ are normalization constants, and $E_{n}$ is the energy of the $n$-th state.

For the evaluation of our experiment, we also need to consider the time evolution during free fall because the distance from the exit of the absorber/scatterer to the $^{10}$B$_{4}$C layer of the detector was $\ell_0=\SI{1.5}{\milli \metre}$ (see Figure~\ref{fig:setup_ILL}).
Suppose that the $n$-th wave function at the exit of the mirror and absorber/scatterer is the initial state $\psi_n(y,t=0)$, then the wave function reaching the $^{10}$B$_{4}$C layer can be computed from~\cite{wave_fun},
\begin{equation}
\psi_n(y,t)=\int K(y,y',t)\psi_n(y',t=0)\,\mathrm{d}y'
\label{eq:psi_t},
\end{equation}
where
\begin{equation}
K(y,y',t)=\sqrt{\frac{m}{2\pi \hbar t}}\exp{\left(\frac{\mathrm{i}}{\hbar}S(y,y',t)\right)}
\label{eq:K}
\end{equation}
and
\begin{equation}
S(y,y',t)=\frac{m}{2t}\left(y'-y-\frac{1}{2}gt^2\right)^2+mgyt-\frac{1}{6}mg^2t^2
\label{eq:S}.
\end{equation}
Here, $t=l_{0}/v$, $v$ is the horizontal velocity of the neutrons.

In order to compare our theoretical prediction based on Equation~\ref{eq:psi_t} with our measurements, we used a model which was given by an ideal function with infinitely high spatial resolution.
However, the fit resulted in an extremely small $p$-value of \num{5.4e-17}, indicating a low statistical probability that the model represents the data well and the necessity to add the effect of the blurring to the model.
Therefore, we used a model given by the following function which was a convolution of the previous model and Gaussian function:
\begin{equation}
f(y)=A  \sum_{n=1}^N\left(\frac{P_n}{\sqrt{2\pi}\sigma}\int_{-3\sigma}^{3\sigma}|\psi_n(y-y_s-y'',t)|^2\, \exp{\left(-\frac{y''^2}{2\sigma^2}\right)}\,\mathrm{d}y''\right)+B
\label{eq:fit_fun_fin_res},
\end{equation}
where $N$ is the maximum number of the quantum state, $A$ is the magnification of the distribution, $B$ is the amount of background noise, $\sigma$ represents the blurring of the distribution, and $P_{n}$ is the relative probability of quantum state $n$ obeying $\sum_{n=1}^N P_{n} = 1$.
In Equation~\ref{eq:fit_fun_fin_res}, the effects of interference terms are ignored, which is reasoned by these terms becoming extremely small on average due to the random phase of each neutron.

This model was then used in the least squares fits to the histogrammized data from our measurement.
We used $A$, $B$, $y_s$, $\sigma$, and $P_{n}$ ($n$ = 1, 2, ..., $N$-1) as free parameters.
$N$ and $d_e$ were varied by hand but kept fixed for each fit.
The true value of the effective gap $d_{e}$ was assumed to lie between \SI{26}{\micro \metre} and \SI{36}{\micro \metre}, considering the measured value of the gap and the effect of the roughness.
We performed fits with the values $d_e=26,\, 31,\,\text{ and }\SI{36}{\micro \metre}$.
For $N$, we initially restricted ourselves to the most probable cases $N=3$ and $N=4$.
Generally, for $N=4$, the probability $P_4$ of the state $n=4$ becomes almost zero (except for the case of $d_e=\SI{26}{\micro\meter}$).
In the latter case, $P_4$ was the largest of all states, which we consider an artifact caused by fixing $d_e$ to a value far from the real one.
Therefore, the number of states was fixed to be $N=3$ in all subsequent fits.

\begin{figure}[htbp]
\begin{center}
\includegraphics[scale=0.45,clip]{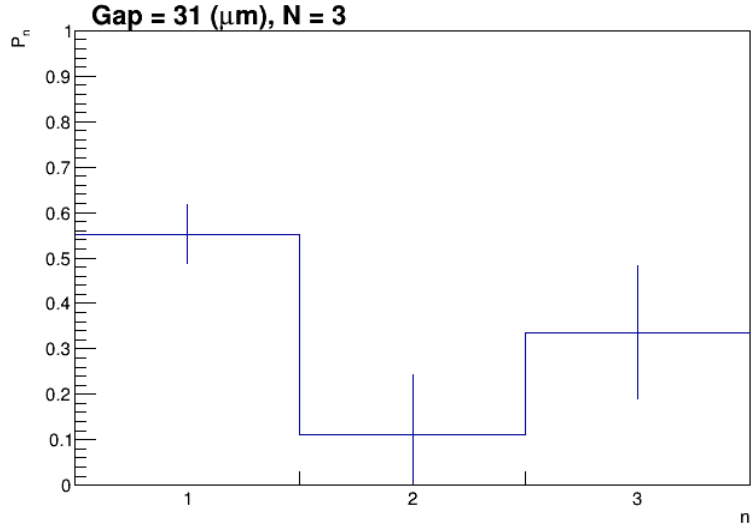}
\caption{Relative probabilities of each quantum state for the fit result shown in Figure~\ref{fig:fit_result_combo}.}
\label{fig:fit_pop_combo}
\end{center}
\end{figure}

\newpage
Figure~\ref{fig:fit_result_combo} and Figure~\ref{fig:fit_pop_combo} show the fit results and the contributing probability distributions for each quantum state using the model $f(y)$ for fixed effective gap of $d_e=\SI{31}{\micro \metre}$.
The best fit with $p$-value 0.70 was obtained for $\sigma=\SI{6.9}{\micro \metre}$, indicating a valid result.
The accuracy of determining the probability of each quantum state was lowered due to this blurring.

The cause of the difference in resolution between cold neutrons (used for the measurements with the Gd grating) and UCNs was considered to be the effect of neutron refraction due to the large surface roughness of the upstream (beam side) of the Si substrate in front of the emulsion layer.
The refraction was negligible at the downstream side (emulsion side) of Si substrate as the surface roughness was polished. 
To evaluate this effect, we measured the height of the roughness $z(x,\,y)$ of the surface of the Si substrate in a \SI{560x560}{\micro \metre} region with an interval of \SI{2}{\micro \metre}, where $z$ is a coordinate value in the direction of the thickness of the substrate, using a 3D Optical Profiler (Zygo NewView 6000).
From the results, we obtained a set of values,
\begin{equation}
g(x,x',y) = \langle(z(x,y)-z(x',y))^2\rangle,
\label{eq:zygo_data}
\end{equation}
as discussed in Reference~\cite{roughness}.
For the estimation of the blur described below, we followed the procedure outlined in Reference~\cite{scatter}. 
Then, the set was transformed to $g'(r)$, where $r=\sqrt{x^{2}-x'^{2}}$ is the absolute distance between two measurement points.
Finally, $g'(r)$ was fitted by the following function,
\begin{equation}
f(r) = 2b^2\left(1-\exp{\left(-\left(\frac{r}{\sqrt{2}w}\right)^2\right)}\right)\,,
\label{eq:fit_roughness}
\end{equation}
where $b$ and $w$ are the standard deviation of the surface roughness and the correlation length, respectively.
From the fit results, we obtained $b=\SI{0.293 \pm 0.006}{\micro \metre}$ and $w=\SI{3.44 \pm 0.20}{\micro \metre}$.
For these values, the classical refraction effect is dominant for UCNs.
Assuming that the neutrons hit the silicon surface at normal incidence, the refraction angle can be computed:
\begin{equation}
\theta = \frac{b}{\sqrt{2}w}\frac{k_s^2}{k_n^2}
\label{eq:refraction},
\end{equation}
Using the values $k_{n}=\SI{0.15}{\per\nano \metre}$ for the wavenumber of the neutrons for a mean velocity $v= \SI[per-mode=symbol]{9.5}{\metre\per\second}$, and $k_{s}=\SI{0.051}{\per\nano \metre}$ for the critical wavenumber of Si, and we arrive at $\theta=\SI{6.9}{\milli \radian}$.
The resulting blur (spread of the waveform) for our Si substrate having a thickness of \SI{0.5}{\milli \metre} is \SI{3.4}{\micro \metre}.
In the case of the resolution measurement using the transmission image of the Gd grating, the typical velocity of cold neutrons used in the measurement was \SI[per-mode=symbol]{1000}{\metre\per\second} and the wavenumber was \SI{16}{\per\nano \metre}; hence, the refraction angle was \SI{0.62}{\micro \radian} and the blur was \SI{0.31}{\nano \metre}.
The magnitude of the latter blur was negligible compared to the resolution of the detector.
Thus, the observed difference of the measured resolution between cold neutrons and UCNs can be explained quantitatively.

In order to estimate the blur in the case of a double-side-polished Si substrate, we measured $b$ and $w$ of the polished surface of its substrate in the same manner as described above. 
As a result, we obtained $\SI{0.60 \pm 0.04}{\nano \metre}$ and $\SI{42 \pm 8}{\micro \metre}$, respectively.
In case $b$ is smaller than the UCN wavelength, the surface scattering will follow the coherent scattering described by the micro-roughness model~\cite{scatter}.
In this model, \SI{6.9}{\percent} of the total UCNs would diffuse and the most likely scattering angle was \SI{150}{\micro \radian} if a non-divergent beam is used under the same conditions as in the UCN experiment described above.
It corresponds to \SI{75}{\nano \metre} for \SI{0.5}{\milli \metre} thick Si.
Since this is small enough compared to the  characteristic scale of $y_{0}$ in Equation~\ref{eq:boundary_3}, it is expected that UCNs in the quantized states can be clearly observed by using the emulsion detector with a double-side-polished Si substrate.

\section{Conclusion}
Different experiments use the system of a quantum bouncer, realized by the quantized states of cold neutrons or UCNs on a flat mirror in the Earth's gravitational field, to search for non-Newtonian interactions.
For this purpose, neutron detectors with a high spatial resolution are used.
In this study, we improved upon a detector so that it could measure the spatial distribution of UCNs with a high spatial resolution using a fine-grained nuclear emulsion.
Two tasks were accomplished with the detector to conduct such measurements.
The first task involved establishing a method to ensure that the high spatial accuracy and resolution can be achieved within a wide region of \SI{65x0.2}{\milli \metre}, for which reference marks were created using electron beam lithography.
The second task involved the establishment of a method for the emulsion to work in vacuum, for which a holder was developed to maintain the volume around the emulsion at atmospheric pressure.

A transmission image of a Gd grating was acquired by cold neutron detection at BL05 of the J-PARC Material and Life Science Experimental Facility in Japan, and the resolution of the improved detector was evaluated by fitting the edge portion of the image with an error function.
The results demonstrate that the intrinsic resolution of the detector was better than \SI{0.56 \pm 0.08}{\micro \metre} for cold neutrons.

A test exposure was conducted to obtain a spatial distribution of UCNs in quantized states on a mirror in the Earth's gravitational field at PF2 of the Institut Laue-Langevin.
The distribution was obtained, fitted with the theoretical curve, and turned out to be reasonable for UCNs in quantized states when we considered a blurring of \SI{6.9}{\micro \metre}.
The blurring was well explained as a result of neutron refraction due to the large surface roughness on the upstream side of the Si substrate.
By using a double-side-polished Si substrate, a resolution of less than \SI{0.56}{\micro \metre} is expected to be achieved for UCNs.

\acknowledgments
We thank T. Naka for providing us fine-grained nuclear emulsion gel and advice regarding its use.
We are grateful to T. Shinohara, T. Samoto, and A. Momose for his help in using the \SI{9.00}{\micro \metre} Gd grating and its electron micrograph.
We also thank P. Geltenbort for constructive advice and T. Brenner for his tremendous support on the experiment at Institut Laue-Langevin.
The sputtering of the converter layer and test experiments for the development of the detector was conducted under the support of common use programs of the Institute for Integrated Radiation and Nuclear Science, Kyoto University.
The reference marks were developed and created at Nagoya University Research Facility for Advanced Science and Technology with the assistance of T. Kato, D. Oshima and M. Sakashita.
This research was supported by JSPS KAKENHI Grant Number 18H05210.
The neutron experiment at the Materials and Life Science Experimental Facility of the J-PARC was performed under user programs (Proposal No. 2017B0336, 2018A0263, and 2019A0227) and S-type project of KEK (Proposal No. 2019S03).
\bibliography{bib}
\bibliographystyle{unsrt}
\end{document}